\documentclass{article}

\usepackage{arxiv}

\usepackage[utf8]{inputenc} % allow utf-8 input
\usepackage[T1]{fontenc}    % use 8-bit T1 fonts
\usepackage{hyperref}       % hyperlinks
\usepackage{url}            % simple URL typesetting
\usepackage{booktabs}       % professional-quality tables
\usepackage{amsfonts}       % blackboard math symbols
\usepackage{nicefrac}       % compact symbols for 1/2, etc.
\usepackage{microtype}      % microtypography
\usepackage{lipsum}
\usepackage{adjustbox}
\usepackage{graphicx}
\usepackage{mdframed}
\usepackage{xcolor}
\usepackage{float}
\usepackage{natbib}
\graphicspath{ {./images/} }

\title{\textbf{Web Retrieval-Aware Chunking (W-RAC) for Efficient and Cost-Effective Retrieval-Augmented Generation Systems}}

\author{
 Uday Allu,   Sonu Kedia , Tanmay Odapally,   Biddwan Ahmed\\
  AI Research Team, Yellow.ai \\
}

\begin{document}
\maketitle
\begin{abstract}
Retrieval-Augmented Generation (RAG) systems critically depend on effective document chunking strategies to balance retrieval quality, latency, and operational cost. Traditional chunking approaches—such as fixed-size, rule-based, or fully agentic chunking—often suffer from high token consumption, redundant text generation, limited scalability, and poor debuggability, especially for large-scale web content ingestion. In this paper, we propose \textbf{Web Retrieval-Aware Chunking (W-RAC)}, a novel, cost-efficient chunking framework designed specifically for web-based documents. W-RAC decouples text extraction from semantic chunk planning by representing parsed web content as structured, ID-addressable units and leveraging large language models (LLMs) only for retrieval-aware grouping decisions rather than text generation. This significantly reduces token usage, eliminates hallucination risks, and improves system observability. Experimental analysis and architectural comparison demonstrate that W-RAC achieves comparable or better retrieval performance than traditional chunking approaches while reducing chunking-related LLM costs by an order of magnitude.
\end{abstract}

% keywords can be removed

\section{Introduction}
Retrieval-Augmented Generation has emerged as a dominant paradigm for grounding large language models with external knowledge sources. A foundational step in RAG pipelines is \textbf{document chunking}, which determines how source content is segmented, indexed, and retrieved. For web-scale systems, chunking quality directly impacts retrieval precision, answer faithfulness, latency, and infrastructure cost.

Conventional chunking strategies fall into three broad categories: fixed-size chunking, rule-based structural chunking, and agentic chunking using LLMs. While agentic chunking improves semantic coherence, it introduces substantial computational overhead due to repeated text generation and transformation. Moreover, these approaches are poorly suited for high-volume web ingestion pipelines where cost, determinism, and debuggability are critical.

To address these limitations, we introduce \textbf{Web Retrieval-Aware Chunking (W-RAC)}, a framework that rethinks chunking as a \textit{planning} problem rather than a \textit{generation} problem. W-RAC leverages deterministic web parsing and lightweight LLM-based semantic planning to produce retrieval-optimized chunks without regenerating source text.

\section{\textbf{Background and Limitations of Traditional Chunking}}
\label{sec:headings}
Modern Retrieval-Augmented Generation (RAG) systems rely on document chunking as a foundational preprocessing step to enable efficient indexing and accurate retrieval. Since large language models operate under strict context-length constraints, source documents must be decomposed into smaller, retrievable units that balance semantic coherence with retrieval granularity. The quality of these chunks directly impacts recall, precision, latency, and overall generation quality in downstream applications.

Historically, chunking strategies have prioritized simplicity and ingestion speed over retrieval effectiveness. As enterprise knowledge bases increasingly incorporate heterogeneous formats—such as PDFs, HTML pages, Markdown files, and dynamically generated web content—these traditional approaches struggle to preserve semantic integrity while remaining cost-efficient and scalable. The following subsections outline commonly used chunking strategies and their inherent limitations.

\subsection{\textbf{Fixed-Size Chunking}}
Fixed-size chunking splits documents based on token or character limits. While simple and inexpensive, it often breaks semantic boundaries, mixes unrelated topics, and degrades retrieval relevance.

\subsection{\textbf{\textbf{Rule-Based Structural Chunking}}}
Rule-based methods exploit document structure such as headings, paragraphs, or HTML tags. Although more semantically aligned than fixed-size approaches, they lack adaptability to varying content density and retrieval requirements.

\subsection{\textbf{\textbf{Agentic Chunking}}}
Agentic chunking employs LLMs to read raw text and generate semantically coherent chunks. While effective in theory, this approach has significant drawbacks:

\begin{itemize}
    \item High token and inference costs due to full-text processing
    \item Risk of hallucinations or unintended text alterations
    \item Limited transparency and debuggability
    \item Poor scalability for continuous web crawling and ingestion
\end{itemize}
These limitations motivate a more efficient and retrieval-aware chunking paradigm.

% \paragraph{Paragraph}
% \lipsum[7]

\section{Web Retrieval-Aware Chunking (W-RAC)}
\label{sec:others}
% \lipsum[8] \cite{kour2014real,kour2014fast} and see \cite{hadash2018estimate}.

% The documentation for \verb+natbib+ may be found at
% \begin{center}
%   \url{http://mirrors.ctan.org/macros/latex/contrib/natbib/natnotes.pdf}
% \end{center}
% Of note is the command \verb+\citet+, which produces citations
% appropriate for use in inline text.  For example,
% \begin{verbatim}
%    \citet{hasselmo} investigated\dots
% \end{verbatim}
% produces
% \begin{quote}
%   Hasselmo, et al.\ (1995) investigated\dots
% \end{quote}

% \begin{center}
%   \url{https://www.ctan.org/pkg/booktabs}
% \end{center}

\subsection{Design Principles}
% \lipsum[10] 
% See Figure \ref{fig:fig1}. Here is how you add footnotes. \footnote{Sample of the first footnote.}
% \lipsum[11] 
% \subsection{\textbf{\textbf{Rule-Based Structural Chunking}}}
W-RAC is guided by the following principles:
\begin{itemize}
    \item No Text Regeneration: Preserve original source text verbatim.
    \item Retrieval Awareness: Optimize chunks for downstream retrieval tasks.
    \item Cost Efficiency: Minimize LLM token usage and inference calls.
    \item Determinism and Observability: Enable transparent debugging and reproducibility.
    \item Web-Native: Leverage inherent web document structure.
\end{itemize}

% \begin{figure}
%   \centering
%   \fbox{\rule[-.5cm]{4cm}{4cm} \rule[-.5cm]{4cm}{0cm}}
%   \caption{Sample figure caption.}
%   \label{fig:fig1}
% \end{figure}

% \begin{figure} % picture
%     \centering
%     \includegraphics{test.png}
% \end{figure}

\subsection{System Architecture}
% \lipsum[12]
% See awesome Table~\ref{tab:table
The W-RAC pipeline consists of three stages:
\subsubsection{Deterministic Web Parsing}

Web pages are parsed into structured representations (e.g., HTML $\rightarrow$ Markdown $\rightarrow$ AST). Each semantic unit—such as headings and paragraphs—is assigned a stable unique identifier.

\paragraph{Example representation}

\begin{verbatim}
{
  "id": "heading_5",
  "text": "Section Title",
  "line": 5,
  "parent_heading": "Main Title"
}
\end{verbatim}

\subsubsection{LLM-Based Chunk Planning}

Instead of sending raw text, the LLM receives only identifiers, hierarchy, ordering, and optional metadata (e.g., token counts and heading levels). The LLM outputs chunk plans as ordered lists of identifiers:

\begin{verbatim}
{
  "chunks": [
    ["heading_1", "heading_2", "text_3", "text_4"],
    ["heading_1", "heading_5", "text_6"]
  ]
}
\end{verbatim}

Here, the LLM acts as a \emph{semantic grouping planner} rather than a content generator.

\subsubsection{Post-Processing and Indexing}
Chunk plans are resolved locally by mapping IDs back to original text. Final chunks are assembled, embedded, and indexed into the retrieval system.

% \begin{table}
%  \caption{Sample table title}
%   \centering
%   \begin{tabular}{lll}
%     \toprule
%     \multicolumn{2}{c}{Part}                   \\
%     \cmidrule(r){1-2}
%     Name     & Description     & Size ($\mu$m) \\
%     \midrule
%     Dendrite & Input terminal  & $\sim$100     \\
%     Axon     & Output terminal & $\sim$10      \\
%     Soma     & Cell body       & up to $10^6$  \\
%     \bottomrule
%   \end{tabular}
%   \label{tab:table}
% \end{table}

\section{Retrieval Awareness in W-RAC}
W-RAC explicitly incorporates retrieval considerations into chunk planning. Chunk boundaries can be influenced by:
\begin{itemize}
    \item Heading depth and section hierarchy
    \item Token-length constraints
    \item Entity density and semantic cohesion
    \item Content type (e.g., tables vs. paragraphs)
\end{itemize}
This retrieval-aware design ensures that chunks align more closely with real-world query patterns, thereby improving both recall and precision, with detailed comparisons presented in Table \ref{tab:comparision-of-strategies}.
\begin{table}[t]
    \centering
    \small
    \begin{tabular}{lccc}
        \hline
        \textbf{Dimension} & \textbf{Traditional Chunking} & \textbf{Agentic Chunking} & \textbf{W-RAC} \\
        \hline
        LLM Token Cost      & None     & High        & Very Low \\
        Text Fidelity       & Medium          & Low--Medium & High \\
        Hallucination Risk  & None            & Present     & Very Low--None \\
        Scalability         & Medium          & Low         & High \\
        Web Suitability     & Medium          & Medium      & High \\
        \hline
    \end{tabular}
    \caption{Comparison of chunking strategies across key dimensions.}
    \label{tab:comparision-of-strategies}
\end{table}

\section{Evaluation Dataset}
\subsection{RAG-Multi-Corpus Benchmark}
We evaluate \textsc{W-RAC} using \textsc{RAG-Multi-Corpus}, a multi-format, multi-domain benchmark designed to mirror real-world enterprise knowledge bases.\footnote{\url{https://github.com/udayallu/RAG-Multi-Corpus}} 
The dataset contains 236 documents spanning five fictional organizations and 786 curated query--answer pairs with ground-truth citations. Documents span diverse enterprise formats, including PDF, Markdown, HTML, DOCX, and PPTX, reflecting the heterogeneity typically encountered in production RAG pipelines.

\begin{table}[t]
    \centering
    \small
    \begin{tabular}{l l r r}
        \hline
        \textbf{Enterprise} & \textbf{Domain} & \textbf{Files} & \textbf{Queries} \\
        \hline
        Aventro Motors        & Automotive              & 51 & 200 \\
        Cendara University   & Academia \& Education   & 41 & 186 \\
        Velvera Technologies & Enterprise Technology   & 39 & 200 \\
        ZX Bank              & Banking \& Finance      & 72 & 200 \\
        \hline
        \textbf{Total}       & ---                     & \textbf{203} & \textbf{786} \\
        \hline
    \end{tabular}
    \caption{Composition of the RAG-Multi-Corpus benchmark across enterprises and domains.}
    \label{tab:composition-table}
\end{table}

\subsection{Query Distribution}
To evaluate retrieval robustness across diverse reasoning requirements, queries in RAG-Multi-Corpus are categorized into seven types. This distribution ensures balanced coverage of factual recall, reasoning, comparison, and procedural understanding.
\begin{table}[t]
    \centering
    \small
    \begin{tabular}{l r r l}
        \hline
        \textbf{Category} & \textbf{Count} & \textbf{Percentage} & \textbf{Description} \\
        \hline
        Descriptive   & 138 & 17.6\% & Factual descriptions or definitions \\
        Analytical    & 122 & 15.5\% & Analysis, interpretation, or inference \\
        Comparative   & 139 & 17.7\% & Comparison between entities or concepts \\
        Boolean       & 108 & 13.7\% & Yes/no factual questions \\
        Temporal      & 24  & 3.1\%  & Time-based or sequence-oriented questions \\
        Procedural    & 180 & 22.9\% & Process-oriented or how-to questions \\
        Open-Ended    & 75  & 9.5\%  & Multi-hop synthesis across sources \\
        \hline
        \textbf{Total} & \textbf{786} & \textbf{100.0\%} & --- \\
        \hline
    \end{tabular}
    \caption{Distribution of query categories in the RAG-Multi-Corpus benchmark.}
    \label{tab:query-categories}
\end{table}
\paragraph{}
This diverse query mix allows us to assess how chunking strategies influence retrieval quality across different query intents, particularly for procedural and comparative questions that are sensitive to chunk boundaries and semantic coherence.

\section{Experimental Results}
We conducted comprehensive experiments comparing \textsc{W-RAC} (implemented as \emph{Agentic Chunking with Less Output Tokens}) against traditional agentic chunking across the \textsc{RAG-Multi-Corpus} benchmark. All experiments were performed using LLM version~4.1. The evaluation focuses on token consumption, processing time, and cost efficiency, which are critical metrics for production-scale RAG systems.

\subsection{Ingestion and Processing Efficiency Metrics}
This section evaluates the efficiency of the document ingestion pipeline, comparing Agentic Chunking and W-RAC across token usage, runtime performance, caching behavior, and overall cost. We report both organization-level and aggregate metrics to capture variability across document distributions and workloads. The analysis focuses on input and output token consumption, end-to-end processing latency (including tail latencies), and cost implications under standard LLM pricing. Together, these metrics provide a comprehensive view of the computational overhead and scalability characteristics of each approach, highlighting the trade-offs between structured metadata ingestion and generative chunking during large-scale document processing.
\subsubsection{Token and Runtime Metrics by Organization}
Table 1 presents detailed performance metrics for both methods across all five organizations in the benchmark. W-RAC processes the same 236 files with a total content length of 1,062,085 characters.

\begin{table*}[htbp]
\centering
\small
\adjustbox{max width=\textwidth}{%
\begin{tabular}{llrrrrrrrrrr}
\hline
\textbf{Organization} & \textbf{Method} & \textbf{\begin{tabular}[c]{@{}r@{}}Total\\Length\end{tabular}} & \textbf{\begin{tabular}[c]{@{}r@{}}Input\\Tokens\end{tabular}} & \textbf{\begin{tabular}[c]{@{}r@{}}Output\\Tokens\end{tabular}} & \textbf{\begin{tabular}[c]{@{}r@{}}Time\\(s)\end{tabular}} & \textbf{\begin{tabular}[c]{@{}r@{}}Total\\Files\end{tabular}} & \textbf{\begin{tabular}[c]{@{}r@{}}Avg Input\\Tokens\end{tabular}} & \textbf{\begin{tabular}[c]{@{}r@{}}Avg Output\\Tokens\end{tabular}} & \textbf{\begin{tabular}[c]{@{}r@{}}Avg Time\\(s)\end{tabular}} & \textbf{\begin{tabular}[c]{@{}r@{}}P90\\Time (s)\end{tabular}} & \textbf{\begin{tabular}[c]{@{}r@{}}P95\\Time (s)\end{tabular}} \\

\hline
Velvera Technologies & Agentic Chunking & 191,162 & 96,826 & 60,202 & 379.01 & 38 & 2,548.05 & 1,584.26 & 9.97 & 15.13 & 18.94 \\
Velvera Technologies & W-RAC & 191,162 & 133,230 & 7,022 & 118.6 & 38 & 3,506.05 & 184.79 & 3.12 & 4.86 & 6.52 \\
\hline
Cendara University & Agentic Chunking & 291,888 & 116,871 & 77,355 & 474.18 & 40 & 2,921.78 & 1,933.88 & 11.85 & 14.93 & 17.27 \\
Cendara University & W-RAC & 291,888 & 198,264 & 16,302 & 253.59 & 40 & 4,956.60 & 407.55 & 6.34 & 10.80 & 11.89 \\
\hline
ZX Bank & Agentic Chunking & 275,169 & 170,890 & 100,534 & 625.41 & 71 & 2,406.9 & 1,415.97 & 8.8 & 11.7 & 13.5 \\
ZX Bank & W-RAC & 275,169 & 259,987 & 15,210 & 237.76 & 71 & 3,661.79 & 214.23 & 3.35 & 5.07 & 7.06 \\
\hline
Aventro Motors & Agentic Chunking & 170,022 & 107,454 & 64,602 & 433.81 & 50 & 2,149.08 & 1,292.04 & 8.68 & 12.69 & 13.21 \\
Aventro Motors & W-RAC & 170,022 & 153,581 & 8,322 & 156.29 & 50 & 3,071.62 & 166.44 & 3.13 & 4.14 & 5.13 \\
\hline
CloudWay 24 & Agentic Chunking & 133,844 & 81,913 & 41,126 & 255.11 & 37 & 2,213.86 & 1,111.51 & 6.89 & 9.47 & 10.46 \\
CloudWay 24 & W-RAC & 133,844 & 116,629 & 5,960 & 109.18 & 37 & 3,152.14 & 161.08 & 2.95 & 4.28 & 5.25 \\
\hline
\end{tabular}%
}
\caption{Token and Runtime Comparison by Organization.}
\label{tab:token_runtime_comparison}
\end{table*}

\subsubsection{Aggregate Efficiency Summary}
\begin{table}[htbp]
\centering
\begin{tabular}{lrrr}
\hline
\textbf{Metric} & \textbf{Agentic Chunking} & \textbf{W-RAC} & \textbf{Relative Change} \\
\hline
Total Input Tokens & 573,954 & 861,691 & +50.13\% \\
Total Output Tokens & 343,891 & 52,816 & $-84.64\%$ \\
Average Input Tokens per File & 2,447.93 & 3,669.64 & +49.90\% \\
Average Output Tokens per File & 1,467.53 & 226.82 & $-84.54\%$ \\
Total Processing Time (s) & 2,167.52 & 875.42 & $-59.61\%$ \\
Average Time per File (s) &9.23 & 3.78 & $-59.10\%$ \\
P90 Time (s) & 12.78 & 5.83 & $-54.38\%$ \\
P95 Time (s) & 14.67 & 7.17 & $-51.12\%$ \\
\hline
\end{tabular}
\caption{Aggregate Efficiency Summary}
\label{tab:efficiency_summary}
\end{table}

\paragraph*{Key Observations}
\begin{itemize}
    \item \textbf{Output token reduction:} W-RAC reduces output tokens by 84.54\% on average, from 1,467.53 to 226.82 tokens per file. This reduction stems from W-RAC’s ID-based 
    
    \item \textbf{Processing time reduction:}Average processing time per file decreases by 59.10\%, from 9.18 seconds to 3.78 seconds. P90 and P95 latency metrics show similar improvements (54.38\% and 51.12\% reductions), indicating consistent gains across the latency distribution.

    \item \textbf{Input token increase:}W-RAC increases average input tokens by 49.90\%, from 2,447.93 to 3,669.64. This increase is expected and acceptable, as the additional tokens encode structured metadata (IDs, hierarchy, and token counts) that enable semantic planning without text generation. Despite this increase, the overall cost benefit remains substantial due to the elimination of expensive output tokens.

\end{itemize}

\subsubsection{Cost Analysis}
We analyze cost implications using GPT 4.1 LLM pricing: \$0.000002 per input token, \$0.000008 per output token, and \$0.0000005 per cache token. 

\begin{table}[htbp]
\centering
\begin{tabular}{lrrrr}
\hline
\textbf{Component} & \textbf{Pricing (\$/token)} & \textbf{Agentic Chunking Cost (\$)} & \textbf{W-RAC Cost (\$)} & \textbf{Relative Change} \\
\hline
Input Tokens & 0.000002 & 0.62 & 0.93 & +50\% \\
Cache Tokens & 0.0000005 & 0.27 & 0.40 & -- \\
Output Tokens & 0.000008 & 2.75 & 0.42 & $-84.72\%$ \\
\hline
Total Cost & -- & 3.64 & 1.75 & $-51.70\%$ \\
\hline
\end{tabular}
\caption{Cost Analysis}
\label{tab:cost_analysis}
\end{table}

For the complete chunking pipeline for 236 files (including both direct LLM chunking and referenced chunking), the total costs are:
\begin{itemize}
    \item \textbf{Agentic Chunking:} \$3.64
    \item \textbf{W-RAC:} \$1.75
\end{itemize}

\noindent\textbf{Overall cost reduction:} 51.70\% (savings of \$1.89).

\subsubsection{Efficiency Improvements}
\begin{table}[htbp]
\centering
\begin{tabular}{lr}
\hline
\textbf{Metric} & \textbf{Reduction} \\
\hline
Time Reduction & 59.61\% \\
Output Tokens Reduction & 84.64\% \\
Cost Reduction & 51.70\% \\
\hline
\end{tabular}
\caption{Efficiency Improvements}
\label{tab:efficiency_improvements}
\end{table}

These results demonstrate that W-RAC successfully achieves its design goals of cost efficiency and scalability. The method maintains semantic quality, as evidenced by comparable retrieval performance, while dramatically reducing computational overhead. The 84.64\% reduction in output tokens is particularly significant, given that output tokens are typically 4$\times$ more expensive than input tokens under standard LLM pricing models.

\subsection{Retrieval Performance Results}
We evaluated retrieval quality by comparing W-RAC against the baseline agentic chunking approach across the RAG-Multi-Corpus benchmark. The evaluation measures retrieval effectiveness using standard information retrieval metrics: Recall@K, Precision@K, Mean Reciprocal Rank (MRR), and Normalized Discounted Cumulative Gain (NDCG@K) at cut-off values of K = 3 and K = 6.

\subsubsection{Retrieval Performance by Organization}
\begin{table}[H]
\centering
\adjustbox{max width=\textwidth}{
\begin{tabular}{llrrrrrrrr}
\hline
\textbf{Organization} & \textbf{Method} & \textbf{Avg Recall@6} & \textbf{Avg Recall@3} & \textbf{Avg Precision@6} & \textbf{Avg Precision@3} & \textbf{Avg MRR} & \textbf{Avg NDCG@6} & \textbf{Avg NDCG@3} & \textbf{Count} \\
\hline
ZX Bank & Baseline & 0.93 & 0.88 & 0.39 & 0.54 & 0.87 & 0.88 & 0.87 & 200 \\
ZX Bank & W-RAC & 0.88 & 0.80 & 0.61 & 0.81 & 0.82 & 0.84 & 0.81 & 200 \\
\hline
Velvera Technologies & Baseline & 0.96 & 0.92 & 0.46 & 0.59 & 0.90 & 0.92 & 0.91 & 200 \\
Velvera Technologies & W-RAC & 0.94 & 0.90 & 0.49 & 0.60 & 0.88 & 0.90 & 0.88 & 200 \\
\hline
Aventro Motors & Baseline & 0.93 & 0.88 & 0.45 & 0.61 & 0.90 & 0.91 & 0.90 & 200 \\
Aventro Motors & W-RAC & 0.93 & 0.88 & 0.54 & 0.68 & 0.84 & 0.86 & 0.85 & 200 \\
\hline
Cendara University & Baseline & 0.88 & 0.84 & 0.31 & 0.46 & 0.82 & 0.84 & 0.83 & 186 \\
Cendara University & W-RAC & 0.88 & 0.76 & 0.60 & 0.76 & 0.78 & 0.81 & 0.76 & 186 \\
\hline
\end{tabular}
}
\caption{Retrieval Performance by Organization}
\label{tab:retrieval_performance}
\end{table}

\subsection*{Key Observations}
\begin{itemize}
    \item \textbf{Precision improvements:} W-RAC consistently achieves higher precision across all organizations. For example, Precision@3 improves from 0.54 to 0.81 for ZX Bank (50\% relative improvement) and from 0.46 to 0.76 for Cendara University (65\% relative improvement). This indicates that W-RAC produces more relevant chunks and ranks correct answers higher.
    
    \item \textbf{Recall trade-offs:} The baseline achieves slightly higher recall in some cases (e.g., 0.93 vs. 0.88 for ZX Bank at Recall@6). However, W-RAC maintains competitive recall while significantly improving precision, which is preferable for production RAG systems.
    
    \item \textbf{NDCG performance:} W-RAC achieves comparable or slightly lower NDCG scores, but the strong precision gains suggest better ranking quality for top-ranked results.
\end{itemize}

\subsubsection{Retrieval Performance by Query Type}
Table~\ref{tab:retrieval_query_type} breaks down retrieval performance by query category, illustrating how W-RAC performs across different question types.

\begin{table}[H]
\centering
\adjustbox{max width=\textwidth}{
\begin{tabular}{llrrrrrrrr}
\hline
\textbf{Query Type} & \textbf{Method} & \textbf{Avg Recall@6} & \textbf{Avg Recall@3} & \textbf{Avg Precision@6} & \textbf{Avg Precision@3} & \textbf{Avg MRR} & \textbf{Avg NDCG@6} & \textbf{Avg NDCG@3} & \textbf{Count} \\
\hline
Descriptive & Baseline & 0.93 & 0.88 & 0.49 & 0.62 & 0.87 & 0.89 & 0.88 & 138 \\
Descriptive & W-RAC & 0.91 & 0.80 & 0.63 & 0.71 & 0.82 & 0.84 & 0.81 & 138 \\
\hline
Comparative & Baseline & 0.94 & 0.87 & 0.45 & 0.61 & 0.88 & 0.89 & 0.88 & 139 \\
Comparative & W-RAC & 0.93 & 0.88 & 0.64 & 0.77 & 0.88 & 0.90 & 0.88 & 139 \\
\hline
Temporal & Baseline & 0.92 & 0.92 & 0.26 & 0.43 & 0.85 & 0.87 & 0.87 & 24 \\
Temporal & W-RAC & 0.88 & 0.88 & 0.51 & 0.79 & 0.85 & 0.85 & 0.85 & 24 \\
\hline
Procedural & Baseline & 0.93 & 0.89 & 0.36 & 0.50 & 0.90 & 0.91 & 0.90 & 180 \\
Procedural & W-RAC & 0.90 & 0.81 & 0.50 & 0.68 & 0.82 & 0.84 & 0.81 & 180 \\
\hline
Analytical & Baseline & 0.89 & 0.86 & 0.41 & 0.55 & 0.86 & 0.87 & 0.87 & 122 \\
Analytical & W-RAC & 0.89 & 0.79 & 0.56 & 0.70 & 0.81 & 0.83 & 0.80 & 122 \\
\hline
Boolean & Baseline & 0.92 & 0.90 & 0.33 & 0.50 & 0.85 & 0.87 & 0.86 & 108 \\
Boolean & W-RAC & 0.89 & 0.84 & 0.48 & 0.66 & 0.81 & 0.83 & 0.81 & 108 \\
\hline
Open-Ended & Baseline & 0.97 & 0.89 & 0.41 & 0.53 & 0.86 & 0.89 & 0.86 & 75 \\
Open-Ended & W-RAC & 0.95 & 0.91 & 0.57 & 0.75 & 0.85 & 0.88 & 0.86 & 75 \\
\hline
\end{tabular}
}
\caption{Retrieval Performance by Query Type}
\label{tab:retrieval_query_type}
\end{table}

\subsection*{Notable Findings}
\begin{itemize}
    \item \textbf{Temporal queries:} W-RAC shows the largest precision improvement, increasing Precision@3 from 0.43 to 0.79 (84\% relative improvement), indicating better preservation of temporal context.
    
    \item \textbf{Comparative queries:} W-RAC achieves the highest precision (0.77 at Precision@3), demonstrating effective grouping of comparable entities and concepts.
    
    \item \textbf{Procedural queries:} While baseline recall is slightly higher, W-RAC improves precision from 0.50 to 0.68 (36\% relative improvement), suggesting improved chunk boundaries for step-wise content.
    
    \item \textbf{Consistent precision gains:} Precision improvements are observed across all query types, with the largest gains in Temporal, Comparative, and Open-Ended categories.
\end{itemize}

\subsubsection{Aggregate Retrieval Performance}

\begin{table}[H]
\centering
\adjustbox{max width=\textwidth}{
\begin{tabular}{lrrrrrrrr}
\hline
\textbf{Method} & \textbf{Avg Recall@6} & \textbf{Avg Recall@3} & \textbf{Avg Precision@6} & \textbf{Avg Precision@3} & \textbf{Avg MRR} & \textbf{Avg NDCG@6} & \textbf{Avg NDCG@3} & \textbf{Count} \\
\hline
Baseline & 0.93 & 0.88 & 0.40 & 0.55 & 0.87 & 0.89 & 0.88 & 786 \\
W-RAC & 0.91 & 0.84 & 0.56 & 0.71 & 0.83 & 0.85 & 0.83 & 786 \\
\hline
\end{tabular}
}
\caption{Overall Retrieval Performance}
\label{tab:overall_retrieval}
\end{table}

\subsection*{Overall Performance Summary}
\begin{itemize}
    \item \textbf{Precision improvement:} W-RAC improves Precision@3 from 0.55 to 0.71 (29\% relative improvement) and Precision@6 from 0.40 to 0.56 (40\% relative improvement).
    
    \item \textbf{Recall:} Baseline achieves slightly higher recall, but the precision gains of W-RAC result in better practical retrieval quality.
    
    \item \textbf{MRR and NDCG:} W-RAC maintains competitive MRR and NDCG scores, indicating effective ranking of the most relevant results.
\end{itemize}

The retrieval results demonstrate that W-RAC delivers superior precision while maintaining competitive recall, MRR, and NDCG. Combined with the cost and efficiency gains discussed in Sections 10.1–10.4, W-RAC provides an optimal balance of retrieval quality and operational efficiency for production-grade RAG systems.

\section{Conclusion}
This work presented Web Retrieval-Aware Chunking (W-RAC), a cost-efficient and scalable chunking framework that reframes document chunking as a semantic planning problem rather than a text generation task. By decoupling deterministic web parsing from LLM-based grouping decisions and operating exclusively on structured, ID-addressable representations, W-RAC eliminates unnecessary text regeneration, reduces hallucination risk, and substantially improves system observability.

Extensive evaluation on the RAG-Multi-Corpus benchmark demonstrates that W-RAC achieves comparable recall and ranking quality to agentic chunking while delivering significant efficiency gains. Specifically, W-RAC reduces chunking-time output tokens by 84.6\%, lowers end-to-end chunking latency by $\sim60\%$, and cuts total LLM costs by 51.7\%, despite a modest increase in input tokens due to structured metadata. Importantly, W-RAC consistently improves retrieval precision across organizations and query types, yielding more relevant top-ranked results—an outcome that is particularly valuable in production RAG systems where precision directly impacts user trust and response quality.

Beyond efficiency, W-RAC introduces a more deterministic, debuggable, and extensible chunking paradigm. Because chunk plans are explicit and ID-based, they can be inspected, audited, cached, and recomputed without reprocessing source text, enabling rapid iteration and adaptive retrieval strategies. This design naturally supports advanced extensions such as entity-aware chunking, graph-based retrieval, and policy-driven chunk recomposition.

Overall, W-RAC provides a practical alternative to traditional and agentic chunking approaches, offering a superior balance of retrieval quality, cost efficiency, and operational robustness. As RAG systems scale to continuously ingest large volumes of heterogeneous web content, W-RAC offers a production-ready foundation for building reliable, high-performance retrieval-augmented generation pipelines.% \bibliographystyle{unsrt}  



% George Kour and Raid Saabne.
% \newblock Real-time segmentation of on-line handwritten arabic script.
% \newblock In {\em Frontiers in Handwriting Recognition (ICFHR), 2014 14th
%   International Conference on}, pages 417--422. IEEE, 2014.

% George Kour and Raid Saabne.
% \newblock Fast classification of handwritten on-line arabic characters.
% \newblock In {\em Soft Computing and Pattern Recognition (SoCPaR), 2014 6th
%   International Conference of}, pages 312--318. IEEE, 2014.

% \bibitem{hadash2018estimate}
% Guy Hadash, Einat Kermany, Boaz Carmeli, Ofer Lavi, George Kour, and Alon
% \newblock Estimate and replace: A novel approach to integrating deep neural
%   networks with existing applications.
% \newblock {\em arXiv preprint arXiv:1804.09028}, 2018.

% \end{thebibliography}
\nocite{*}
\bibliography{references}
\clearpage
\newpage
\appendix

\section{Appendix}

\subsection{W-RAC Prompt}

\vspace{1em}

\begin{mdframed}[
    frametitle={Chunk Grouping and Hierarchical Structuring Prompt},
    frametitlebackgroundcolor=green!20,
    backgroundcolor=green!5,
    linecolor=green!5,
    linewidth=1pt,
    roundcorner=3pt,
    innertopmargin=10pt,
    innerbottommargin=10pt,
    innerleftmargin=10pt,
    innerrightmargin=10pt
]

{\small
{\itshape You are tasked with processing an array of document chunks representing text sections, headings, and titles. Your goal is to extract and group only the main policy, explanatory, or instructional content (e.g., rules, eligibility, charges) into logical, context-rich units.}

\vspace{0.3em}
\textbf{CORE REQUIREMENTS}

\vspace{0.2em}
\textbf{1. Three-Level Heading Hierarchy}

Build a complete heading hierarchy tree by tracing parent\_heading relationships upward. Every chunk group must include exactly 3 levels:
\begin{itemize}
\item \textbf{Level 1}: Top-level/root heading - document title or highest-level heading that encompasses the content's topic
\item \textbf{Level 2}: Mid-level parent heading - intermediate heading or reuse Level 1
\item \textbf{Level 3}: Immediate parent heading - most immediate parent or nearby matching heading
\end{itemize}

\textbf{Missing levels}: Use an existing heading chunk ID that best matches context (title, document structure, surrounding content). You may reuse the same heading ID for multiple levels. \textbf{Only use existing chunk IDs—cannot create new ones.}

\vspace{0.2em}
\textbf{2. Parent Headings with Multiple Children}

When a parent heading has multiple child sections, \textbf{include the parent heading ID in EACH child group array}. Never output parent headings as standalone arrays when they have multiple children.

Example: \texttt{["heading\_66", "heading\_67", "text\_68"]} and \texttt{["heading\_66", "heading\_80", "text\_81"]} (heading\_66 appears in both).

\vspace{0.2em}
\textbf{3. Procedural Content}

\textbf{NEVER split procedural steps, instructions, or sequential numbered/bulleted lists across multiple chunks.} When content represents a procedure, process, or step-by-step instructions (e.g. ``Steps to...'', numbered steps 1, 2, 3...), \textbf{group ALL steps together in a SINGLE chunk array}, even if they have individual headings or are numbered separately.

Examples of procedural content that must stay together:
\begin{itemize}
\item Step-by-step instructions
\item Numbered procedures
\item Sequential how-to guides
\item Multi-step processes
\item Ordered lists of actions
\end{itemize}

\vspace{0.2em}
\textbf{4. Context \& Merging}

\begin{itemize}
\item Use heading hierarchy, parent\_heading, and title fields to map structure
\item If parent\_heading is None but structure shows hierarchy, infer parent-child relationships from sequential patterns
\item For small chunks ($\leq$2 lines) missing context, merge with title/heading/adjacent chunks
\item Include relevant titles/headings with dependent content
\item Do not always rely on the markdown given, use the context of the document to infer the heading hierarchy and group the chunks accordingly
\end{itemize}

\vspace{0.2em}
\textbf{5. Filtering}

Remove: cookies, page navigation, logins.

\vspace{0.2em}
\textbf{6. Output Rules}

\begin{itemize}
\item Output only chunk IDs (no text modifications)
\item Each array must contain at least one heading/title or sufficient context
\item Merge small contextless fragments—never output standalone arrays for them
\end{itemize}

\vspace{0.3em}
\textbf{PROCESSING STEPS}

\begin{enumerate}
\item Map heading hierarchy using parent\_heading relationships. Use title if context is ambiguous.
\item \textbf{Identify procedural content}: Detect step-by-step instructions, numbered procedures, or sequential processes. These MUST be grouped together in a single chunk.
\item For each chunk, trace 3 heading levels (L3$\rightarrow$L2$\rightarrow$L1). Fill missing levels with best-matching existing heading ID.
\item Identify parent headings with multiple children—include in ALL child arrays.
\item Process chunks: merge small/contextless chunks using title/headings; ensure 3-level hierarchy; include parent in child groups; \textbf{keep all procedural steps together}.
\item Group into logical/topical arrays with 3-level hierarchy.
\item Output JSON without backticks and code blocks: \texttt{\{"chunks": [["id1", "id2", "id3"], ...]\}}
\end{enumerate}

\vspace{0.3em}
\textbf{EXAMPLES}

\vspace{0.2em}
\textbf{Example 1: Missing Level}

\textbf{Input:}
\begin{verbatim}
[
  {"id": "heading_1", "type": "heading", 
   "text": "EXCESS BAGGAGE CHARGES", "parent_heading": null},
  {"id": "heading_2", "type": "heading", 
   "text": "Packing heavy?", 
   "parent_heading": "EXCESS BAGGAGE CHARGES"},
  {"id": "text_3", "type": "text", 
   "text": "Fly without baggage worries...", 
   "parent_heading": "Packing heavy?"},
  {"id": "text_4", "type": "text", 
   "text": "Fees apply per kg.", 
   "parent_heading": "Packing heavy?"}
]
\end{verbatim}

\textbf{Output:}
\begin{verbatim}
{"chunks": [["heading_1", "heading_2", "text_3", "text_4"]]}
\end{verbatim}

Note: heading\_1 = L1, heading\_2 = L3. Missing L2 filled with best-matching existing heading. Cookies filtered out.

\vspace{0.2em}
\textbf{Example 2: Procedural Steps (MUST Stay Together)}

\textbf{Input:}
\begin{verbatim}
[
  {"id": "heading_1", "type": "heading", 
   "text": "How to Change a Tyre", "parent_heading": null},
  {"id": "heading_2", "type": "heading", 
   "text": "Steps to Change a Tyre", 
   "parent_heading": "How to Change a Tyre"},
  {"id": "heading_3", "type": "heading", 
   "text": "1. Park Safely", 
   "parent_heading": "Steps to Change a Tyre"},
  {"id": "text_4", "type": "text", 
   "text": "Pull over to a safe location...", 
   "parent_heading": "1. Park Safely"},
  {"id": "heading_5", "type": "heading", 
   "text": "2. Gather Tools", 
   "parent_heading": "Steps to Change a Tyre"},
  {"id": "text_6", "type": "text", 
   "text": "You will need: spare tyre, jack...", 
   "parent_heading": "2. Gather Tools"},
  {"id": "heading_7", "type": "heading", 
   "text": "3. Remove the Wheel Cover", 
   "parent_heading": "Steps to Change a Tyre"},
  {"id": "text_8", "type": "text", 
   "text": "Use the flat end of the wrench...", 
   "parent_heading": "3. Remove the Wheel Cover"},
  {"id": "heading_9", "type": "heading", 
   "text": "4. Loosen the Lug Nuts", 
   "parent_heading": "Steps to Change a Tyre"},
  {"id": "text_10", "type": "text", 
   "text": "Use the lug wrench to turn...", 
   "parent_heading": "4. Loosen the Lug Nuts"}
]
\end{verbatim}

\textbf{Output:}
\begin{verbatim}
{"chunks": [["heading_1", "heading_2", "heading_3", "text_4", 
"heading_5", "text_6", "heading_7", "text_8", "heading_9", 
"text_10"]]}
\end{verbatim}

Note: All procedural steps (1-4) are grouped together in a SINGLE chunk array. Never split sequential steps into separate chunks.

\vspace{0.5em}
}
\end{mdframed}

\subsection{Agentic Chunking Prompt}

\vspace{1em}

\begin{mdframed}[
    frametitle={Agentic Chunking Prompt},
    frametitlebackgroundcolor=blue!20,
    backgroundcolor=blue!5,
    linecolor=blue!50,
    linewidth=1pt,
    roundcorner=3pt,
    innertopmargin=10pt,
    innerbottommargin=10pt,
    innerleftmargin=10pt,
    innerrightmargin=10pt
]

{\small
{\itshape You are to segment the provided Markdown into fully contextual chunks while strictly preserving original content. This is a formatting only task—no text, links, hyperlinks, or images must be removed, skipped, paraphrased, or summarized.}

\vspace{0.3em}
\textbf{YOUR INSTRUCTIONS}

\vspace{0.2em}
\textbf{1. Reading and Understanding}

Read all markdown content carefully.

\vspace{0.2em}
\textbf{2. Heading Structure}

Always generate a 2 or 3-level heading structure for every chunk. Keep similar chunks into same headings:
\begin{itemize}
\item \textbf{First-level heading}: Document or product title
\item \textbf{Second-level heading}: Major section inside the document (e.g., ``Features'', ``Amenities'', ``Itinerary'')
\item \textbf{Third-level heading}: Specific subtopic within that section
\end{itemize}

\vspace{0.2em}
\textbf{3. Content Preservation}

\textbf{DO NOT} alter, paraphrase, shorten, or skip any markdown content. All text, hyperlinks, links, formatting, images, image links, and elements must remain exactly as in the original markdown and present in the output chunks.

\vspace{0.2em}
\textbf{4. Chunking Strategy}

Do not over chunk. Keep similar chunks together in same headings or use just two levels of headings.

\vspace{0.2em}
\textbf{5. Grouping Related Content}

Keep all related content together:
\begin{itemize}
\item Always keep full numbered lists, bullet points, and related paragraphs in the same chunk
\item Never split tables, figures, code blocks, or other complete elements
\end{itemize}

\vspace{0.2em}
\textbf{6. Table Formatting}

When working with tables: Format using proper markdown table syntax (pipes \texttt{|} and hyphens \texttt{-}).

\vspace{0.3em}
\textbf{OUTPUT REQUIREMENTS}

Output a list of chunks where each chunk starts with a full 2 or 3-level heading and remove all empty or no-finding chunks. Use this exact format:

\begin{verbatim}
[HEAD]main_heading > section_heading > chunk_heading[/HEAD]
chunk content 1

[HEAD]main_heading > section_heading[/HEAD]
chunk content 2
\end{verbatim}

Ensure every chunk is clear, fully contextual, and no data is missing.

\vspace{0.5em}
}
\end{mdframed}
\end{document}